\documentclass[manuscript]{acmart}
\usepackage{wrapfig}
\usepackage[subtle]{savetrees}
\usepackage{amsmath,amsfonts}
\newcommand{\nobjects}{n}
\newcommand{\capacity}{c}
\newcommand{\stepToLocNo}{x}
\newcommand{\stepToLocOptNo}{\stepToLocNo^\star}
\newcommand{\stepToLocTakenNo}{\bar\stepToLocNo}
\newcommand{\stepToLoc}[1]{\stepToLocNo(#1)}
\newcommand{\stepToLocOpt}[1]{\stepToLocOptNo(#1)}
\newcommand{\stepToLocTaken}[1]{\stepToLocTakenNo(#1)}
\newcommand{\locToStep}[1]{\stepToLocNo^{-1}(#1)}
\newcommand{\natno}[1]{\{1,\ldots,#1\}}
\newcommand{\natnoZero}[1]{\{0,\ldots,#1\}}
\newcommand{\objectLocations}{\mathcal P}
\newcommand{\receptacleLocations}{\mathcal D}
\newcommand{\indicatorFunction}[1]{{\boldsymbol 1}_{#1}}
\newcommand{\indicatorFunctionTwo}[2]{\indicatorFunction{#1}(#2)}
\newcommand{\distanceNo}{d}
\newcommand{\distance}[2]{\distanceNo(#1,#2)}
\newcommand{\receptacleNo}{f}
\newcommand{\receptacle}[1]{\receptacleNo(#1)}

\newcommand{\RRplus}[1]{\mathbb R_+^{#1}}
\usepackage{titlesec}
\titlespacing*{\section}{0pt}{0.3\baselineskip}{0.3\baselineskip}
\titlespacing*{\subsection}{0pt}{0.1\baselineskip}{0.1\baselineskip}
\AtBeginDocument{%
  \providecommand\BibTeX{{%
    \normalfont B\kern-0.5em{\scshape i\kern-0.25em b}\kern-0.8em\TeX}}}

\setcopyright{acmcopyright}
\copyrightyear{2021}
\acmYear{2021}
\acmDOI{10.1145/1122445}

\acmConference[College Station '21]{College Station '21: ACM Conference on Intelligent User Interfaces}{April 13--17, 2021}{College Station, Texas}
\acmBooktitle{College Station '21: Conference on Intelligent User Interfaces,
 April 13--17, 2021, College Station, Texas}
\acmPrice{15.00}
\acmISBN{978-1-4503-XXXX-X/18/06}

\begin{document}

\title{Optimal Assistance for Object-Rearrangement Tasks in Augmented Reality}

\newcommand{\KTC}[1]{{\color{blue}{KTC: #1}}}
\newcommand{\BN}[1]{{\color{red}{BN: #1}}}
\newcommand{\RD}[1]{{\color{magenta}{RD: #1}}}
\author{Benjamin Newman}
\authornote{Work done at Facebook Reality Labs Research}
\email{newmanba@cmu.edu}
\affiliation{%
  \institution{Carnegie Mellon University}
  \streetaddress{5000 Forbes Ave.}
  \city{Pittsburgh}
  \state{PA}
   \country{USA}}

\author{Kevin Carlberg}
\email{carlberg@fb.com}
\affiliation{%
 \institution{Facebook Reality Labs Research}
 \streetaddress{9845 Willows Road NE}
 \city{Redmond}
 \state{Washington}
 \country{USA}}

\author{Ruta Desai}
\email{rutadesai@fb.com}
\affiliation{%
 \institution{Facebook Reality Labs Research}
 \streetaddress{9845 Willows Road NE}
 \city{Redmond}
 \state{Washington}
 \country{USA}}

\renewcommand{\shortauthors}{Newman, et al.}

\begin{abstract}
% Recent advances in AI open up the opportunity for building next-generation assistants that can reason about embodied human interactions with the world. The goal of this work is to provide optimal assistance to a user completing a household chore in a simulated augmented reality environment and to study the effect of such assistance on the user. Specifically, we study the effect of assistant type on (1) task performance, (2) cognitive burden, and (3) sense of agency in natural navigation and object-manipulation tasks of different difficulty levels.
Augmented-reality (AR) glasses---which will have access to real-time, high-fidelity data regarding a user's environment via onboard sensors, as well as an ability to seamlessly display real-time information to the user---present a unique opportunity to provide users with assistance in completing quotidian tasks. Many such tasks---such as house cleaning, packing for a trip, or organizing a living space---can be characterized as \textit{object-rearrangement tasks} defined by users navigating through an environment, picking up objects, and placing them in different locations. We introduce a novel framework for computing and displaying AR assistance that consists of (1) associating an optimal action sequence with the policy of an \textit{embodied agent} and (2) presenting this optimal action sequence to the user as \textit{suggestion notifications} in the AR system's heads-up display. The embodied agent comprises a `hybrid' between the AR system and the user, in that it has the observation space (i.e., sensor measurements) of the AR system and the action space (i.e., task-execution actions) of the user; its policy is learned by minimizing the the time to complete the task. In this initial study, we assume that the AR system's observations include a map of the environment and real-time localization of the objects and user within that map. These modeling choices allow us to formalize the problem of computing AR assistance for any object-rearrangement task as a planning problem that reduces to solving a capacitated vehicle-routing problem (CVRP), which is a variant of the classical traveling salesman problem (TSP) in combinatorial optimization. In addition, we introduce a novel AR simulator that can enable web-based evaluation of AR-like assistance and associated large-scale data collection; our approach is based on the Habitat \cite{savva2019habitat} simulator for embodied artificial intelligence (AI). Finally, we perform a study that evaluates how users respond to the AR assistance generated by the proposed framework on a specific quotidian object-rearrangement task---house cleaning---using our proposed AR simulator. We perform the study at scale using Amazon Mechanical Turk (AMT)~\cite{AMT} and collect data using psiTurk \cite{gureckis2016psiturk}. In particular, we study the effect of our proposed AR assistance on users' task performance and sense of agency over a range of task difficulties. Our results indicate that providing users with the proposed form of AR assistance improves the user’s overall performance. Additionally, we show that while users report a negative impact to their agency, that they may prefer this when presented with a system that aides them in task completion.
\end{abstract}

\begin{CCSXML}
<ccs2012>
   <concept>
       <concept_id>10010147.10010178.10010199</concept_id>
       <concept_desc>Computing methodologies~Planning and scheduling</concept_desc>
       <concept_significance>500</concept_significance>
       </concept>
   <concept>
       <concept_id>10003120.10003121.10003124.10010392</concept_id>
       <concept_desc>Human-centered computing~Mixed / augmented reality</concept_desc>
       <concept_significance>500</concept_significance>
       </concept>
 </ccs2012>
\end{CCSXML}

\ccsdesc[500]{Computing methodologies~Planning and scheduling}
\ccsdesc[500]{Human-centered computing~Mixed / augmented reality}

\keywords{digital assistance, vehicle routing problem, 3D simulator, crowdsourcing, augmented reality}

\begin{teaserfigure}
\begin{center}
  \includegraphics[width=0.8\textwidth]{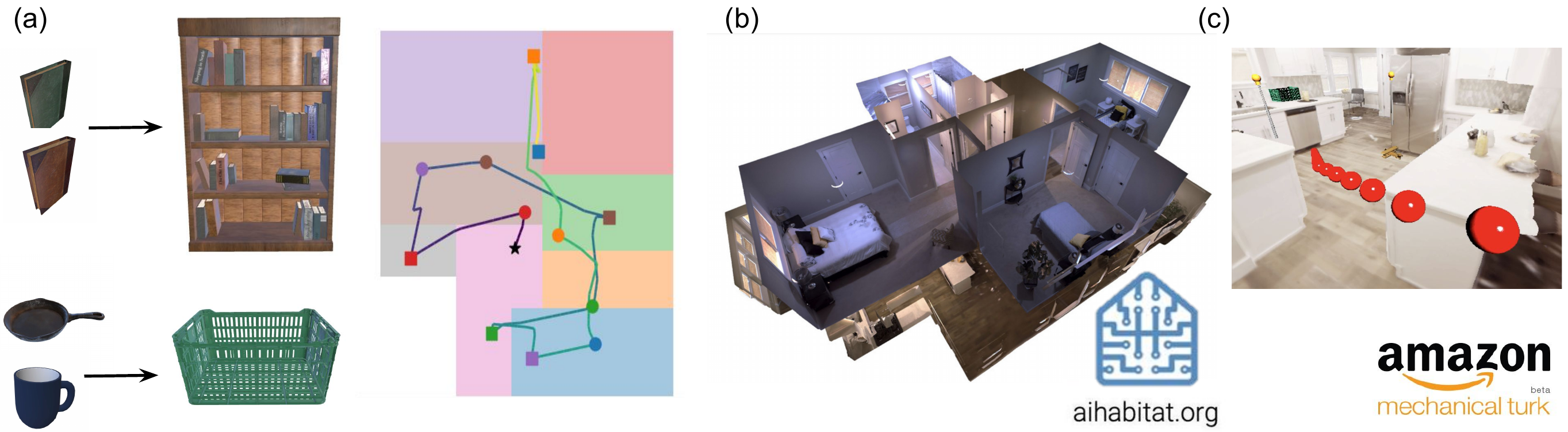}
  \caption{(a)~We present a novel framework for AR assistance in object-rearrangement tasks such as house cleaning that leverages an embodied user-AR `hybrid' agent and a capacitated vehicle routing problem (CVRP) formulation to compute and display the optimal action sequence for the task to the user in AR.~(b)~We also introduce a novel AR simulator based on Habitat~\cite{savva2019habitat} that can enable large-scale web-based evaluation of AR-like assistance (Image courtesy~\cite{straub2019replica}).~(c)~We deploy our framework on Amazon Mechanical Turk~\cite{AMT} to study the effect of our proposed AR assistance on users’ task performance and sense of agency over a range of task difficulties.}
  \Description{Overview of contributions}
  \label{fig:teaser}
\end{center}
\end{teaserfigure}

\maketitle

\section{Introduction}\label{sec:intro}
% TLDR: three main contributions: (1) formulating object-rearrangement as VRP by making assumptions of what the AR system can do, (2) simulated AR environment using Habitat, (3) study to assess \#1 using \#2 on AMT with psiTurk

Compared with current personal computing devices, always-on AR devices~(1)~have access to a much larger volume and more diverse set of sensor data and~(2)~are able to display real-time information to the user in a much lower friction manner \cite{jonker2020}. This exposes the exciting potential for such AR devices to provide users with continual, contextually relevant assistance toward achieving their personalized goals. Consequently, assistive AR systems have been used quite extensively in specialized applications such as maintenance and manufacturing~\cite{palmarini2018systematic, egger2020augmented}, education~\cite{ibanez2018augmented}, tourism~\cite{yung2019new}, and surgery~\cite{vavra2017recent} to name a few. However, AR devices hold the promise of providing assistance to users on a much broader and less specialized class of commonly occurring quotidian activities. For example, if advances in AI can enable AR devices to reason about the structure and state of quotidian tasks such as cooking, cleaning, or organizing, then this ability could be leveraged to lend assistance to users in performing these tasks, ideally leading to improved task performance, reduced physical and cognitive effort, and preserved sense of agency. One may thus envision such AR assistance as providing ``\emph{superpowers for everyday tasks}''. In this work, our goal is to develop a framework for such AR assistance applicable to an important class of everyday tasks---those involving object rearrangement---and to evaluate its value to users at scale. 

Making progress towards this objective of pervasive AR assistance is challenging for myriad reasons. First, there has been limited work towards formalizing the problem of computing and displaying AR assistance that can lead to improved task performance, reduced effort, and preserved sense of agency for users; while this has been investigated for specialized tasks such as AR-assisted assembly~\cite{yang2019influences, tang2003comparative}, it has not yet been pursued for a broad class of quotidian tasks.  Second, no widely available consumer AR device currently exists; as such there is no large AR user base or associated infrastructure that can support the evaluation of AR assistance on users at scale. The field of human--robot interaction (HRI) faces a similar challenge; to overcome it, they have leveraged web-based studies that ask users to react to videos of humans and robots interacting~\cite{hoffman2019evaluating, malle2016robot}. This reliance on a third-person perspective lacks immersion and limits the types of interactions that can be studied; further, no analogous approach to AR assistance would be viable as the AR assistance is not directly observable from third parties. Finally, how real users will respond to AR assistance in everyday tasks---in particular, how it affects their task performance, effort, and sense of agency---remains an open question~\cite{berry2009evaluating, albert2013measuring}.

To address the first challenge in the context of object-rearrangement tasks, we formalize the problem of computing and displaying AR assistance by (1) associating an optimal action sequence with the policy of an \textit{embodied agent} and (2) presenting this optimal action sequence to the user as \textit{suggestion notifications} in the AR system's heads-up display. In our formulation, the embodied agent comprises a user--AR-system `hybrid' in that it has the observation space (i.e., sensor measurements) of the AR system and the action space (i.e., task-execution actions) of the user, and its policy is learned by minimizing the time to complete the task. 
%This single-agent formulation eliminates the need to model the multi-agent dynamics between the human and the AR-system, and it effectively assumes that the user will directly follow the AR-system's suggestions. 
In this initial study, we assume that the AR-system has full observability of the environment, which includes a map and real-time localization of the objects and user within that map. These modeling choices allow us to formalize the problem of computing AR assistance for any object-rearrangement task as a planning task that reduces to solving a capacitated vehicle-routing problem (CVRP) \cite{dantzig1959truck,golden2008vehicle} from combinatorial optimization.
%, a \textit{depot} corresponding to the initial location of the user, \textit{pickup locations} corresponding to the locations of the misplaced objects, \textit{delivery locations} corresponding to the receptacle locations for each object, \textit{transportation costs} corresponding to the geodesic distance between locations, and \textit{capacity constraints} of two. 
Because the optimal action sequence comprises a sequence of location visits along shortest paths, we present this action sequence by displaying the next shortest path to the user in the form of world-locked digital breadcrumbs in the heads-up display. If the user ignores the AR-system's suggestion notifications and deviates from the optimal action sequence by visiting an alternative pickup or delivery location, we replan on the fly.

To address the second challenge, we propose a novel AR simulator that can enable large-scale web-based evaluation of AR assistance and associated data collection. The simulator is based on Habitat \cite{savva2019habitat} for embodied AI and satisfies the key criteria we have in an AR simulator:  (1) it support the observations and actions of the proposed embodied-agent policy,
    (2) it emulates a first-person view through an AR device, including an ability to display suggestion-notifications in a heads-up display (HUD) in the form of digital objects and information, 
    (3) it enables a user in the loop to autonomously perform the task-execution actions, and
    (4) it is deployable on the web at scale via integration with Amazon Mechanical Turk (AMT) and supports data collection related to task performance and sense of agency via psiTurk \cite{gureckis2016psiturk} integration.

To address the third challenge, we define house cleaning as a specific object-rearrangement task, implement the task and the proposed CVRP-based assistance using OR-Tools \cite{OR-Tools} in the proposed AR simulator, and evaluate it at scale using AMT. We collect user data across a range of task difficulties and types of AR assistance in order to evaluate how the proposed form of AR assistance affects users' task performance, effort, and sense of agency. We find that by following the optimal assistance, users are able to decrease their total distance traveled though this comes at a cost of feeling less in control over their own actions. This cost may be one users are willing to pay, however, as we also find that users report preferring the optimal assistance to a system that does not provide them with the optimal solution. Additionally, we find that users are not consistent in their willingness to follow assistance.

In summary, our contributions are:~(1)~a novel framework for computing and displaying AR assistance for object-rearrangement tasks that employs a `hybrid' single agent (i.e., the user--AR-system) and CVRP formulation,~(2)~a novel AR simulator that can enable large-scale web-based evaluation of AR-like assistance, and~(3)~a large-scale web-based study that assesses how users respond to the proposed form of AR assistance in a house-cleaning task over a range of task difficulties. To the best of our knowledge, this is the first at-scale study of AR assistance for quotidian tasks. 
We envision our second contribution as being useful beyond the domain of AR assistance, as it can provide a framework for at-scale user-in-the-loop evaluation of different kinds of digital assistance. Future work will entail developing and assessing perception-based learned policies that assume lower fidelity of the embodied agent's observations and considering multi-agent formulations that incorporate models of the user's behavior.

% Possible contributions:
% 1. We present a framework to study AR assistance in everyday tasks at scale: As a part of the framework, we model users as embodied agents and formulate an everyday household chore of cleaning/re-arranging a house as a vehicle-routing problem. This formulation enables us to compute corresponding optimal task execution that can be offered to the users as assistance. We also leverage a 3D simulation engine called Habitat and deploy it on MTurk to create AR like experiences at scale. 
% 2. Using our framework, we study the effect of optimal task assistance in AR on user's agency and task performance over a range of task difficulties. Our results indicate that providing users with optimal AR assistance improves the user’s overall performance without compromising their sense of agency.

\section{Related Work}
We now review related work that relates to our three key contributions. Namely, we review existing work in AR assistance in Sec.\ \ref{sec:ARassistance}, simulation frameworks for training and evaluating embodied-agent policies in Sec.\ \ref{sec:simulationFrameworks}, and assessing users' response to digital assistance in Sec.\ \ref{sec:usersresponse}.

\subsection{AR assistance}\label{sec:ARassistance}

To date, most work on employing AR devices to assist users has focused on displaying \textit{predefined information overlays} that can be useful in 
completing a prescribed task; the information overlays are often spatially registered to objects or locations relevant to the task. For example, researchers have investigated approaches wherein the AR device overlays information or instructions on parts to be assembled or maintained \cite{thomas1992augmented,tang2003comparative}, retrieves and displays maintenance documentation using object recognition \cite{didier2005amra},  overlaying medical imaging data on patients in real time \cite{bajura1992merging}, or provides location and activity-based, world-locked information to the users during indoor navigation~\cite{mulloni2011handheld}.

Critically, none of the above approaches are truly robust for complex multi-step tasks, as they do not leverage the device's on-board sensors to infer the current task state; as such, they are unable to provide the user with up-to-date assistance toward optimal task completion or properly adapt when users deviate from the system's suggested steps. To fill this gap, planning-based\footnote{Planning refers to the process of computing an agent's policy using a model for the transition dynamics of the environment, i.e., a model that predicts the effect of the agent's actions on the environment states~\cite{sutton2018reinforcement}.} approaches that track task state in real time and update assistance accordingly have been proposed for a variety of AR applications such as robot tele-operation~\cite{tzafestas2001teleplanning, fang2014novel}, assistive surgery~\cite{bourquain2002hepavision2}, assembly and manufacturing~\cite{abramovici2017context}, and even an quotidian task of sandwich assembly~\cite{hu2020interactive}. In particular, Abramovici et al.~\cite{abramovici2017context} enable AR assistance for collaborative manufacturing using a complex back-end infrastructure where smart devices broadcast their status to a centralized planner with a predefined task dependency graph. 

Instead, our proposed approach for AR assistance applicable to complex object-rearrangement tasks described in Sec.\ \ref{sec:arassistance} leverages only the sensor measurements of the AR device, and executes a policy whose state is informed only by these sensor measurements, thus enabling up-to-date assistance toward optimal task completion. In this initial study, we assume the device has `full environment knowledge' via its sensors and can support mapping and localization of the user and objects \cite{livemaps}. This allows us to compute the policy using a planner based on a capacitated vehicle routing problem. Perhaps the most closely related work to ours is the planning-based cognitive assistant developed by Hu et al.~\cite{hu2020interactive}, which employs a perception module driven by computer vision to identify task state in a sandwich-assembly task and re-plan using a finite state machine. 
In contrast to this work, we consider a much broader category of object-rearrangement tasks and consider planners based on a general capacitated vehicle routing problem as described in Sec.~\ref{sec:CVRP}. Further, our embodied AI formulation enables straightforward extensions that go beyond the strict environment-knowledge requirements of planning to learning policies that can employ perception-based partial environment observations.

 \subsection{Embodied-agent simulation frameworks}\label{sec:simulationFrameworks}
% TLDR: our use of simulation is unique: to emulate AR experience and to evaluate at scale
The robotics and embodied AI communities have leveraged simulation frameworks to train and evaluate embodied-agent policies in lieu of expensive real-world experimentation infrastructure ~\cite{savva2019habitat, ai2thor, koenig2004design, erickson2019assistive}. Most efforts consider fully autonomous agents (e.g. robots) whose actions directly change the physical state of the environment. 

Our AR-assistance application is fundamentally different from these, in that actions that can modify the physical state of the environment belong to the human agent. The AR device (our parallel to the autonomous system) can only influence the physical state of the environment indirectly via suggestion-notification actions displayed to the human. Thus, for AR assistance, any realistic \textit{evaluation} of the how the embodied-agent policy affects task performance must involve a human in the loop---or a high-fidelity model of the human's behavior. Evaluating how assistance affects important qualitative aspects of the user experience (e.g., effort, sense of agency) necessitates collecting data from real users at scale. On the other hand, \textit{training} of the embodied-agent policy can be done in a variety of ways that need not rely on a human in the loop; Sec.\ \ref{sec:arassistance} describes in detail our proposed formulation for computing the policy that does not require a human in the loop---only a planner coupled to a simulator that can compute the geodesic distances between locations and track task state---although future work will consider human-in-the-loop training loop analogously to Refs.~\cite{li2016dialogue, zhang2019leveraging}. Thus, for learning assistive AR policies, rather different demands are placed on the environments used for \textit{evaluation} and \textit{training}; the only strong compatibility requirement is that the evaluation environment must support the states and actions associated with the trained policy to deploy it in real time. Here, we focus on the evaluation environment.

Arguably the closest related work is in the field of human--robot interaction (HRI), where researchers are interested in evaluating users' response to robotic assistance~\cite{brooks2018brave}. In this field, the two most common evaluation environments comprise either (1) in-person studies where users are asked to respond to robotic assistance~\cite{dragan2013legibility}, and (2) web-based studies where users are asked to view third-person videos or photos of humans and robots interacting and react accordingly~\cite{hoffman2019evaluating, malle2016robot}. The former category suffers from lack of scalability, while the latter category lacks immersion due to its reliance on a third-person perspective. 

Thus, we believe that our proposed evaluation framework---which is based on an AR simulator and is described in Sec.\ \ref{sec:arsimulator}---combines the attractive attributes of each of these paradigms, as it enables scalable, first-person interaction of a user with an embodied assistant and thus we hypothesize can capture more realistic user experiences---and thus yield higher quality user-experience data---at scale. 

\subsection{Evaluating user response to digital assistance}\label{sec:usersresponse}
It is generally challenging to characterize the user experience for digital assistance owing to a plethora of factors involved and owing to the difficulty of measuring some of these factors with high fidelity~\cite{albert2013measuring}. Past studies on evaluating users' response to digital assistance have focused on evaluating factors such as usability, utility, effectiveness, and agency~\cite{berry2009evaluating}. 
In our study, we focus primarily on evaluating effectiveness as measured by the user's task performance and sense of agency. We focus on task performance as it is easy to measure and directly addresses the key value proposition of AR assistance for complex tasks. We consider agency due to its central role in human--computer interaction (HCI) research and its relatively uncharacterized role in immersive AR-like assistive scenarios.

The sense of agency refers to the feeling of being in the driver's seat when it comes to selecting one's actions~\cite{moore2016sense}. HCI research has long recognized the sense of agency as a key factor in characterizing how people experience interactions with technology~\cite{limerick2014experience,moore2016sense, shneiderman2010designing}. In particular, one of the eight classic rules of interface design places emphasis on designing interfaces that support the user's sense of agency~\cite{shneiderman2010designing}. Further, the sense of agency may also influence a user's acceptance of technology~\cite{baronas1988restoring, le2018agency}. Despite the importance of agency, it has received limited attention in the domain of AR. It is important to address this topic early in developing novel AR technologies, as enabling user's sense of agency is indeed challenging in assistive and immersive systems; for example, previous research has found a reduction in sense of agency with increase in automation~\cite{berberian2012automation}. 

The study presented in Sec.\ \ref{sec:study} evaluates task performance and the user's sense of agency on the proposed cleaning-house task.
Akin to other studies on AR assistance that have shown improved task performance for instance in assembly tasks~\cite{yang2019influences, tang2003comparative}, we show faster task completion times with the use of the proposed AR assistance. Further, we study the effectiveness over a range of task-difficulty and assistance-fidelity levels. 
Our study also suggests that while reports of user agency may be negatively affected by increased assistance in this scenario, this may be a cost participants are willing to pay in order to realize improved task performance; these are promising results, and can serve as a baseline for alternative interfaces for displaying the computed AR assistance to the user.

% Key criteria: task performance, effort, and sense of agency.
% \begin{itemize}
%     \item See this paper on "assessing effectiveness of AR instructions in an assembly task" \cite{tang2003comparative}. They assess "task performance
% and perceived mental workload".
% \end{itemize}
% TL;DR: Limited knowledge about how users respond to assistance that manipulates their sense of physical environment directly.

\section{AR-Assistance Model}\label{sec:arassistance}

The goal of our AR assistance model is to formalize the problem of computing and displaying AR assistance for object-rearrangement tasks.  We achieve this by adopting the perspetive of embodied AI and (1) associate an optimal action sequence with the policy of an embodied agent, and (2) present this optimal action sequence to the user as suggestion notifications in the AR system's heads-up display. To particularize this setup, we must define the embodied agent and associated partially observable Markov decision process (POMDP) ingredients: the states, observations, actions, and rewards characterizing the environment, agent, and task.

\subsection{Embodied AI formulation}

We begin by defining the objective (and associated reward) of the task as moving each object in question from its initial position to its final desired position in as little time as possible. Minimal time to task completion is not only an intuitive choice for an objective and reward, it also draws inspiration from the long history of AI assistants for task and time management~\cite{myers2007intelligent}.

Regarding the choice of embodied agent, one could adopt a fully multiagent perspective and 
consider the user and AR system to be independent but cooperating agents with the shared aforementioned objective but with different observation and action spaces. In this case, one could learn the AR system's policy and present its action sequence to the user as suggestion notifications. Unfortunately, this approach introduces significant challenges, as the user is included in the AR system's environment and thus any simulation-based learning algorithm for the AR-system's policy would require modeling the user's policy acting on an observation space augmented by the AR-system's displayed information \cite{bucsoniu2010multi}. Instead, we simplify this setup and consider a single-agent formulation with an agent comprising a `hybrid' between the AR system and the user. Namely, it has the observation space (i.e., sensor measurements) of the AR system and the actions (i.e., physical task-execution actions) of the user; in the case of object-rearrangement tasks, the latter corresponds to navigation and object pick/place actions. In this case, learning- or planning-based approaches for computing the embodied agent's policy require modeling only the physical dynamics of object rearrangement.

As mentioned above, the embodied-agent's observations correspond to those of the AR system. While many current AR head-mounted devices are equipped with only RGB video, future devices are likely to be equipped with much more high-fidelity observations. Indeed, it is likely that---at least for familiar environments---the AR device will have access to a complete map and will be able to perform localization of the user and objects \cite{livemaps}. As such, for this initial study, we assume that the AR device can has complete observations that include a map of the environment, the current position of all objects in question, and the position of the user. Thus, the observations and state coincide in this case, and the POMDP reduces to an MDP with deterministic transition dynamics, exposing the use for a deterministic planner as described in Sec.\ \ref{sec:CVRP}.

Finally, we assume that the user can carry only two objects at once, the AR device has knowledge of the desired final location of all objects in question (i.e., the `goal state' of the environment), and the AR device can calculate the shortest path between any two points on the map. These assumptions allow us to compute the policy of the embodied agent by a planner that solves a capacitated vehicle routing problem (CVRP), which we describe in Sec.\ \ref{sec:CVRP} below.
We remark that future work will relax the above assumptions and consider multiagent formulations, partial observations, and model-free learning of embodied-agent policies.

\subsection{Object-rearrangement planner: capacitated vehicle routing problem}\label{sec:CVRP}

We now formulate the (single-vehicle) CVRP \cite{dantzig1959truck,golden2008vehicle} for object-rearrangement tasks. We assume that we are rearranging $\nobjects$ objects such that each object has an initial location and final (desired) location, thus yielding $2\nobjects+1$ total locations of interest (including the initial position of the user). Given any arbitrary enumeration of these locations with the zeroth location corresponding to the user's initial position (i.e., the \textit{depot}), we decompose these locations into \textit{pickup locations} $\objectLocations\subset\natno{2\nobjects}$ associating with the objects' current locations and \textit{dropoff locations} $\receptacleLocations\subset\natno{2\nobjects}$ associated with the final locations such that $|\objectLocations| = |\receptacleLocations| = \nobjects$, $\objectLocations\cap\receptacleLocations = \emptyset$, and $\objectLocations\cup\receptacleLocations = \natno{2\nobjects}$. We assume that \textit{transportation costs} can be calculated from an operator $\distanceNo:\natnoZero{2\nobjects}\times \natnoZero{2\nobjects}\rightarrow\RRplus{}$ that calculates the geodesic distance between any two locations and satisfies $\distance{i}{i} = 0$, $i\in\natnoZero{2\nobjects}$. If any two pickup or dropoff locations $i$ and $j$ coincide, we treat them as separate locations with zero separating distance such that $\distance{i}{j}=0$. 
This assumes the user's time to complete the task is proportional to their total path length.
We assume the user has a \textit{capacity constraint} of $\capacity\in\mathbb N$ (which we set to be $\capacity = 2$ in the experiments, which assumes a user can carry one object per hand). We introduce a delivery operator $\receptacleNo:\objectLocations\rightarrow\receptacleLocations$ that maps each pickup location to its corresponding dropoff location. We associate any solution to the problem with an invertible operator $\stepToLocNo:\natnoZero{2\nobjects}\rightarrow\natnoZero{2\nobjects}$ that maps the step number to location index; note that invertibility of this operator assumes that each location is visited exactly once and a location visit is associated with the pickup or dropoff of the appropriate object. Assuming the user has already executed $m(\leq 2\nobjects)$ steps of the task by visiting locations $\stepToLocTaken{0},\ldots,\stepToLocTaken{m}$ with $\stepToLocTaken{0}=0$ and the task initialized at $m=0$, the planner defines the optimal sequence of location visits $(\stepToLocOpt{0},\ldots,\stepToLocOpt{2\nobjects})$ as the solution to the combinatorial optimization problem
\begin{align}\label{eq:CVRP}
\begin{split}
\underset{(\stepToLoc{0},\ldots,\stepToLoc{2\nobjects})}{\text{minimize}}\ &\sum_{i=1}^{2\nobjects}\distance{\stepToLoc{i}}{\stepToLoc{i-1}}\\
\text{subject to}\ &\sum_{i=1}^j\indicatorFunctionTwo{\objectLocations}{\stepToLoc{i}} - \indicatorFunctionTwo{\receptacleLocations}{\stepToLoc{i}}\leq\capacity,\quad \forall j\in\natno{2\nobjects}\\
&\locToStep{\receptacle{i}} > \locToStep{i},\quad \forall i\in\objectLocations\\
& \stepToLoc{i} = \stepToLocTaken{i},\quad \forall i\in\natnoZero{m},
\end{split}
\end{align}
where $\indicatorFunction{A}$ is the indicator function that evaluates to 1 if its argument is in the set $A$ and evaluates to zero otherwise.

The objective function in problem \eqref{eq:CVRP} corresponds to the transportation costs (i.e., total distance traveled); the first set of constraints corresponds to the capacity constraints; the second set of constraints ensures that each object's pickup location is visited before its dropoff location; the third set of constraints enforces that the solution is computed from the current state of the task at step $m$. Given the solution $(\stepToLocOpt{0},\ldots,\stepToLocOpt{2\nobjects})$ to problem \eqref{eq:CVRP}, at step $p(\geq m)$ of the task, the AR device provides assistance by displaying to the user the shortest path between locations $\stepToLocOpt{p}(=\stepToLocTaken{p})$ and $\stepToLocOpt{p+1}$ in the form of world-locked digital breadcrumbs in the heads-up display. If the user `violates' assistance and instead visits a feasible alternative location $\stepToLocTaken{p+1}(\neq\stepToLocOpt{p+1})$, then the system replans by solving problem \eqref{eq:CVRP} with $m\leftarrow p+1$ and resumes. We emphasize that replanning is essential to ensure robustness to realistic user behavior in complex multi-step object-rearrangement tasks.
In practice, we solve planning problem \eqref{eq:CVRP} using OR-Tools \cite{OR-Tools}.

\section{AR simulator and deployment at-scale}\label{sec:arsimulator}

To evaluate how the proposed AR-assistance model described in Section \ref{sec:arassistance} affects the user experience, an AR simulator must
    (1) support the observations and actions of the embodied-agent policy,
    (2) emulate the first-person view through an AR device, including support of the display of suggestion-notifications in the form of digital objects and information in a heads-up display (HUD), 
    (3) enable a user in the loop to autonomously perform the task-execution actions, and
    (4) be deployable on the web at scale and support data collection related to task performance and sense of agency.

To satisfy the above criteria, we build our AR simulator upon 
Habitat~\cite{savva2019habitat}. 
We make this choice because Habitat satisfies all four of these criteria. First, it supports a suite of embodied-agent observations that mimic the on-board sensors of an AR device (e.g., RGBD video, compass) including those relevant to our current formulation (i.e., mapping and localization of objects and the user). Second, it can display world-locked digital objects and other information that can mimic suggestion notifications in an AR system's HUD, thus supporting the actions of the embodied-agent policy. Third, it supports
3D environments based on real-world reconstructions (e.g., the Replica dataset~\cite{straub2019replica}) rather than synthetically generated environments. Fourth, with modifications, it enables users to perform the task-execution actions required for object-rearrangement tasks. In particular, it natively supports navigation, and we implemented a rudimentary pick/place action as described below. 
%\RD{the new habitat api supports pick and place actions. See: \url{https://colab.research.google.com/github/facebookresearch/habitat-lab/blob/master/examples/tutorials/colabs/Habitat_Interactive_Tasks.ipynb#scrollTo=S426fIt1oDgS}} 
Fifth, Habitat supports the WebGL JavaScript API that enables web-based deployment at scale.
Finally, we note that Habitat supports reinforcement-learning training algorithms (e.g., PPO) \cite{wijmans2019dd} that we can employ to train embodied-agent policies in future studies that assume reduced observation fidelity (e.g., assume only RGBD video); this obviates the need to employ different platforms for policy training and evaluation in future work.

\begin{figure*}[t]
    \centering
    \includegraphics[width=0.7\textwidth]{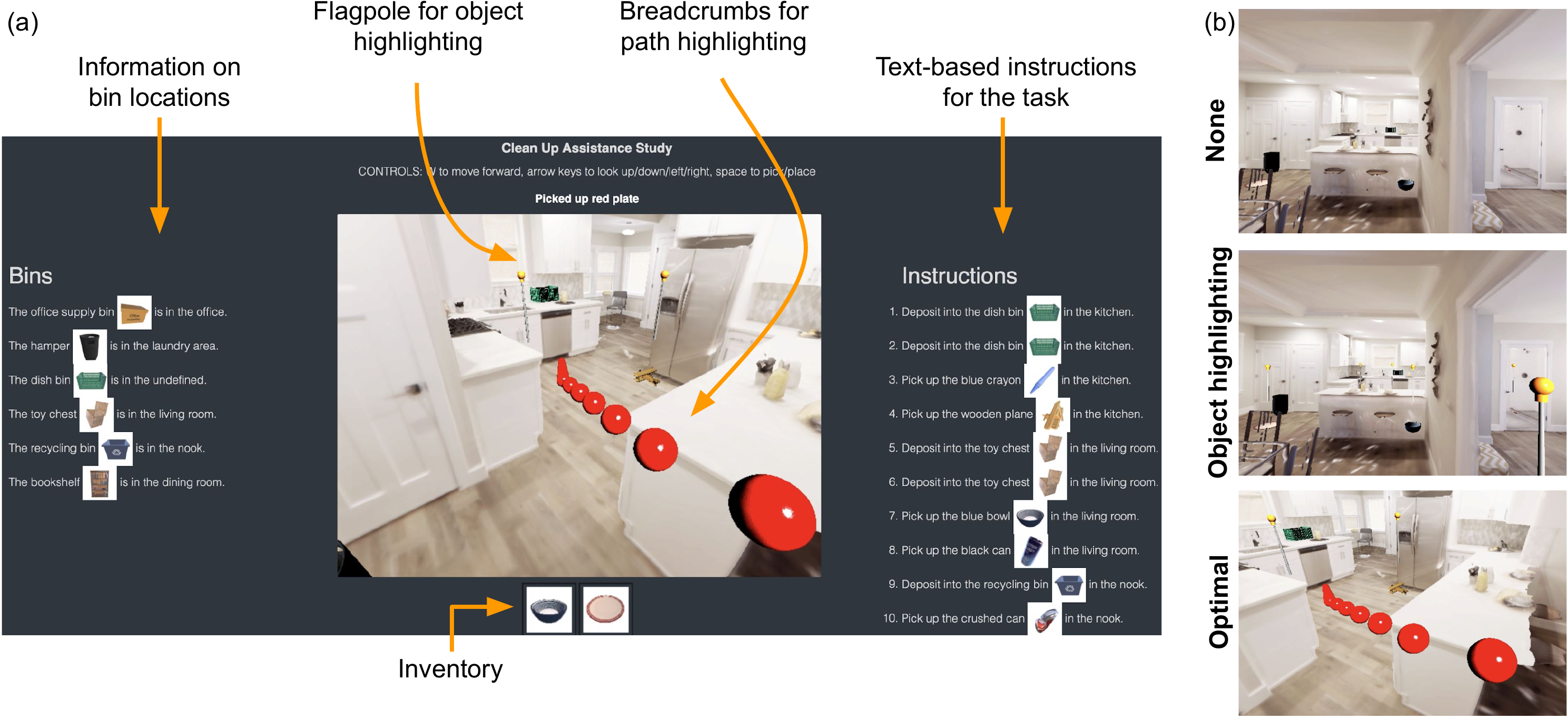}
    \caption{(a)~AR simulator interface with various elements of assistance in the designed virtual HUD.~(b)~Three assistance conditions from the study are shown. Optimal Assistance condition uses both object highlighting via flagpoles and path highlighting via breadcrumbs.}
    \label{fig:assistance_levels}
    \vspace{-0.15in}
\end{figure*}

We make several modifications to Habitat to generate our AR simulator. First, we implement a virtual HUD to mimic the first-person view through an AR device (see Fig.~\ref{fig:assistance_levels}(a)); the HUD supports the display of information relevant to the proposed AR assistance, as well as the ability to display world-locked digital objects in the environment.
A message bar on the top of the HUD alerts users of important interactions, including when the user places an item successfully, or when a user attempts an infeasible action (e.g., attempting to pick/place an item when they are at an excessive distance from a pickup or dropoff location). 
Second, we introduce a `virtual knapsack' of capacity two that represents the user's current inventory; we display the current contents of this backpack as the user's inventory in the virtual HUD.
Third, we introduce a rudimentary object pick/place action that either (1) picks up an object and places it in the virtual knapsack if the user is within a specified distance of a pickup location and the knapsack is not full, or (2) places an object in inventory in its dropoff location if the user is within a specified distance of the dropoff location and the associated object is in inventory. Fourth, we integrate Habitat with the OR-Tools planner to support real-time replanning as described in Section \ref{sec:CVRP}.

By leveraging Habitat's existing WebGL JavaScript API, we can deploy our AR simulator at scale on AMT and collect data related to task performance and sense of agency by leveraging psiTurk \cite{gureckis2016psiturk}. Appendix \ref{sec:system_implementation} describes the details of this setup. 

% \KTC{Use the same arguments from the intro: (1) support of 3D environments based on real-world reconstructions rather than synthetically generated environments, (2) support of embodied-agent observations that mimic the on-board sensors of an AR device (e.g., RGBD video, compass, localization, mapping, object tracking), (3) support of the task-execution actions required for object-rearrangement tasks (i.e., navigation, picking up and placing objects), (4) ability to display world-locked digital objects and other information that can mimic suggestion notifications in an AR system's heads-up display, and (5) support of the WebGL JavaScript API that enables web-based deployment at scale using, e.g., Amazon Mechanical Turk (AMT)}

% 1. Challenge of collecting human data at scale. 
% i. Where we fit on spectrum low-fidelty/high-scale to high-fidelty/low-scale.
% ii. Understanding how this framework can be used to approach high-fidelity or be used in conjunction with high-fidelity/low-scale experiment. 
% Show results from the survey responses on realism use in real world 

% \subsection{Web Deployment on AMT}
% \subsubsection{PsiTurk}
% \subsubsection{Interface}

\section{Study setup}\label{sec:study}

The overarching goal of our study is to evaluate
how users respond to the AR assistance generated by the proposed framework on a specific quotidian object-rearrangement task---house cleaning---using our proposed AR simulator. Sec.\ \ref{sec:task} describes the house-cleaning task. Sec.\ \ref{sec:studyConditions} describes the two key variables that we varied during the study. Sec.\ \ref{sec:metrics} describes the metrics that we employed to evaluate the user experience. Sec.\ \ref{sec:hitoverview} provides an overview of the Human Intelligence Task (HIT) characterizing our study.

\subsection{House-cleaning task}\label{sec:task}
The main house-cleaning task is explained to participants through the following prompt: 
\begin{quote}
In this HIT, you are a guest at a short-term rental. You have been staying at the house for the past several days and checkout time is fast approaching. You must clean up this house according to the host’s instructions in as little time as possible. In order to avoid a late fee, you must navigate through the house to pick up the misplaced items and place them in the appropriate bins before checkout. For instance, you will be asked to place socks in a laundry hamper, books in a bookshelf, dishes in a dish bin, etc... You will be performing the task in a 3D virtual environment using keyboard controls, described later.
\end{quote}

The participant is then placed in a virtual environment within the AR simulator described in Sec.\ \ref{sec:arsimulator} and must complete the task using the keyboard navigation and pick/place actions; they can leverage any AR assistance we provide in the HUD. We consider six semantic object categories and employ a specific bin for each category: dishes (dish bin), toys (toy box), books (bookshelf), laundry (laundry hamper), office supplies (office-supply box), recycling (recycling bin). Fig.~\ref{fig:teaser}(a) illustrates some of the objects and bins used in the study. We note that this formulation leads to a capacitated vehicle routing problem as described in Sec.\ \ref{sec:CVRP}, where $\nobjects$ denotes the number objects to be cleaned up, and delivery locations are repeated when more than one object is associated with the same bin. Appendix \ref{sec:houseGeneration} describes how the bins, objects, and starting location are determined for a given experiment.

\subsection{Study conditions: assistance fidelity and task difficulty}\label{sec:studyConditions}

We vary two key variables across experiments: \textit{assistance fidelity}, which represents how much assistance the AR simulator provides to the user toward efficient task completion; and \textit{task difficulty} as measured by the number of objects that must be cleaned up. We hypothesize that these two variables will be the key drivers of the user experience and task performance. We now describe in more detail how we control these variables.

\subsubsection{Assistance Fidelity}

We consider three different levels of assistance fidelity: no assistance, object-highlighting assistance, and optimal assistance. Each assistance level is characterized by both a world-locked digital-object component and a text-information component; see Fig.~\ref{fig:assistance_levels}. In all conditions, text-information assistance includes a list of bins including their picture and semantic location.

   \textit{No assistance (None)}. Participants receive no assistance from the system in this condition, which serves as our control. The egocentric frame contains only the scene, rendered objects and rendered bins; there are no additional visual cues. The text-information assistance provides a (randomized) list of objects the participant must reorganize in order to complete the task. Each item in this list contains the name of the object, a picture of the object, and the bin in which it should be placed. Once a participant picks up an item, the text corresponding to the selected item is crossed out and move to the bottom of the list.
   
\textit{Object-highlighting assistance}. This form of assistance is designed to provide assistance to the participant under the assumption that the AR device knows the location of the objects and bins salient to the house-cleaning task and can highlight them. Such assistance---which does not rely on knowledge of traversable paths in the environment nor a real-time planner---would be especially helpful in situations where certain items are obstructed from view or are difficult to spot. This form of assistance enables participants to understand the rough locations of all objects at once and form a plan themselves. We implement this form of assistance by placing a digital flagpole over each object as depicted in Fig.~\ref{fig:assistance_levels}; the corresponding text-information assistance includes all of the information from the No-assistance condition but with the addition of the name of the room in which the object can be found. Again, this list is randomized per participant and list items are crossed off as they are completed. 
    
\begin{figure*}[t]
    \centering
    \includegraphics[width=0.85\textwidth]{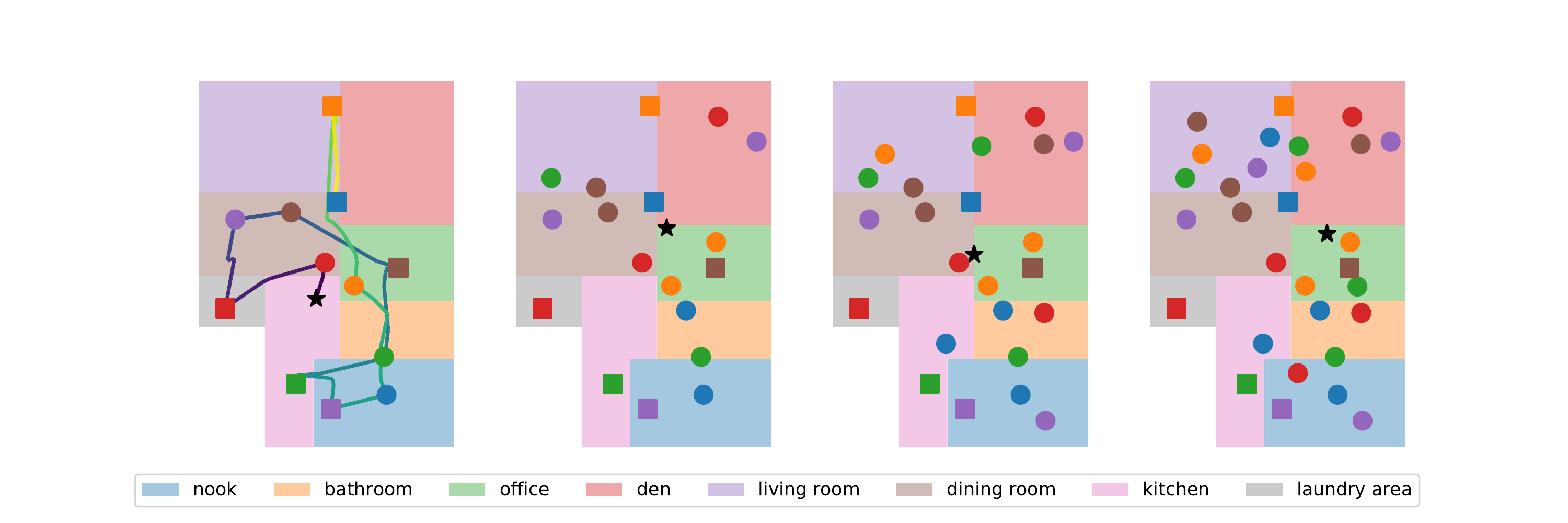}
    \caption{The four task-difficulty levels (increasing in difficulty going from left to right), with the optimal path (corresponding to the solution of the associated CVRP) shown in the leftmost panel. Colored polygons represent approximate bounding boxes of the apartment's floor plan, including the boundaries for each rooms; circular markers denote objects; square markers denote bins; marker color denotes semantic object category (e.g., books); the star represents the initial position.}
    \label{fig:optimal_solution}
\end{figure*}

    \textit{Optimal assistance}. Optimal assistance corresponds to the form of AR assistance proposed in Sec.\ \ref{sec:arassistance}. A solution to this problem for the lowest difficulty setting is shown in Fig.~\ref{fig:optimal_solution}. 
    To display this information to the participant, we display the next segment of the optimal path in the egocentric frame as a trail of digital breadcrumbs, which we set to red spheres. After the user executes a feasible pick/place action, we display the next segment of the optimal path (which may involve replanning as described in Sec.\ \ref{sec:arassistance}. To prevent the participant from losing their orientation with respect to these start and end positions of the optimal path segment, the $z$-coordinate of the path's breadcrumbs start at the participant’s chest level and end at the floor level (see Fig.~\ref{fig:assistance_levels}). In contrast to other assistance conditions, the text-information assistance is now ordered according to the optimal path: each step lists the action the participant should take, the object they should perform it on, a picture of this object, and the object’s location. Additionally, each step is numbered in order to emphasize the importance of the list’s order. As before, items are crossed off the list as they are completed.

\subsubsection{Task Difficulty}

We consider four levels of task difficulty, where we define difficulty in terms of the number of misplaced objects that must be cleaned up. To control for certain objects being easier to reorganize (due to visual salience or bin location), the number of each object type remained fixed for each difficulty setting. A single difficulty level can thus be defined by the ratio of the number of objects to the number of bins. We use four task-difficulty levels, where this ratio corresponds to 1:1 (6 total objects), 2:1 (12 total objects), 3:1 (18 total objects), and 4:1 (24 total objects), respectively. 
Appendix \ref{sec:houseGeneration} describes our approach for generating the bin, object, and initial user locations for any of these conditions.
 Fig.~\ref{fig:optimal_solution} shows the bin locations as squares in each difficulty setting, with each color representing a specific semantic object category.

\subsection{Metrics}\label{sec:metrics} To determine the effect of assistance fidelity and task difficulty on participants' task performance and experience, we collect an array of objective and subjective participant-response data.

\subsubsection{Objective metrics}
We employ four metrics to evaluate task performance:~(1)~\textit{Normalized Deviations}: the number of deviations from the optimal location ordering (accounting for replanning) normalized by the total number of possible deviations,~(2)~\textit{Inverse Path Length (IPL)}: the ratio of the minimal possible path length for to the sequence of location visits taken by the participant\footnote{That is, $\sum_{i=1}^{2n}\distance{\stepToLocTaken{i}}{\stepToLocTaken{i-1}}$ in the notation of Sec.\ \ref{sec:CVRP}.} to the total distance traveled by the participant,~(3)~\textit{Task Distance}: the total distance traveled by the participant, and~(4)~\textit{Task Completion Time}: the ratio of the total time taken by a participant to complete the task to the time taken by a participant to complete the fly-through familiarization phase; this ratio accounts for any system-dependent latency that may affect completion time.

\subsubsection{Subjective metrics}\label{sec:subjective}
We employ subjective metrics that are measured using a five-point Likert scale, which focus on the following two categories\footnote{These questions only comprise a subset of the questions that we asked in the original study. We limit our analysis here to this subset due to space constraints; see the Appendix~\ref{sec:fullsubjective} for the full list.}:

\begin{enumerate}
    \item \textit{Agency}, which is defined as the feeling of being in control~\cite{moore2016sense}.
    \begin{itemize}
        \item[] ``I am in charge of deciding what step I complete next during the house cleaning task'' (\textit{Control what to do})
        %\item[]``I am in charge of deciding where I go next during the house cleaning task''
        \item[]``I am responsible for the speed at which I completed the task'' (\textit{Control of speed})
        \item[]``I feel that I need to follow the suggestions given to me by the system'' (\textit{Need to follow})
        \item[] ``I prefer that the system show me what to do next rather than figure it out myself during the house cleaning task'' (\textit{Prefer to show})
        %\item[] ``I prefer that the system show me where to go next rather than figure it out myself during the house cleaning task''
    \end{itemize}
    \item \textit{Utility and Usability}, which is aimed at measuring usefulness, user-friendliness, and acceptability; it is inspired by the System Usability Scale (SUS)~\cite{bangor2008empirical}.
    \begin{itemize}
        %\item[] ``I found the house cleaning task more difficult to complete than the training task''
        %\item[] ``I thought the system was easy to use'',
        %\item[]``I would imagine that most people would learn to use the help provided by the system very quickly''
        %\item[]``I would like to use the help provided to me by system during the house cleaning task frequently in real life house cleaning scenarios''
        \item[]``The assistance provided to me by the system during the house cleaning task helped me complete the task faster than if I had used the help provided to me during the training task'' (\textit{Usable})
        \item[] ``I found the help given to me by the system to be useful'' (\textit{Useful})
        %\item[]``The help provided to me by the system during the house cleaning task was sufficient in order to accomplish the task''.
    \end{itemize}
\end{enumerate}

\subsection{HIT overview}\label{sec:hitoverview}

%For a complete breakdown of how many participants were in each condition, see Table \ref{tab:participants}.

%\begin{figure*}[ht]
%    \centering
%    \includegraphics[width=0.95\textwidth]{figs/kondo_phases.png}
%    \caption{The five phases of our experiment. Phases 1 through 3 are used to familiarize participants to the task. Phase 4 is the experimental phase, where participants are randomly split into one of the 12 different conditions. Finally, participants are taken to the final phase: the post-experiment survey.}
%    \label{fig:trial_timeline}
%\end{figure*}

The HIT for the study consists of four phases:~(1)~task setup,~(2)~house familiarization and navigation-controls training,~(3)~cleaning-task execution, and~(4)~survey. In the first phase, the participant is presented with the house-cleaning-scenario prompt described in Sec.\ \ref{sec:task}. In the second phase, the participant is familiarized with the 3D layout of the short-term rental house using a pre-recorded fly-through video that displays information regarding room names and bins locations. To acquaint the participant with the keyboard controls, they are given a simple training task of finding and picking up a single object and placing it in its appropriate bin without any imposed time limit or incentive. The third phase corresponds to executing the main cleaning task described in Sec.\ \ref{sec:task} with specified task-difficulty and assistance-fidelity levels. In the final phase, the participant is given a survey that collects demographic information (e.g., age, gender), responses to the subjective questions described in Sec.\ \ref{sec:subjective}, and a free-form response to capture anything else relevant to their experience. Except for the actual task in the third phase of the HIT---which varied across participants depending on the study conditions described in Sec.\ \ref{sec:studyConditions}---all other phases were consistent across participants.

\section{User study}

We conducted a $3\times 4$ between-subjects user study ($n=447$ participants, male$=265$, female$=177$, other$=1$, no answer$=4$; three assistance-fidelity levels; four task-difficulty levels) on AMT using our AR simulator setup described in Sec.~\ref{sec:arsimulator}. Participants were compensated $\$7.50$ for completing the 25-minute-long study. Appendix~\ref{sec:studyparticipants} reports details on the number of participants in each condition. We first present and evaluate the hypotheses related to different assistance fidelity in easiest task difficulty condition using objective and subjective metrics defined in Sec.~\ref{sec:study}. We then show the interaction effects between assistance and task difficulty.  

\subsection{Effects of varying assistance fidelity}\label{sec:assistance}
We now conduct a study that varies the assistance fidelity and fixes task difficulty to the easiest level (i.e., 6 total objects).

\subsubsection{Hypotheses.}
With regards to varying the assistance fidelity, we had the following hypotheses:
\begin{itemize}
    \item[\textbf{H1}] Participants will follow optimal assistance when presented with it.
    \item[\textbf{H2}] Participants will have higher task performance when presented with optimal assistance.
    % \item[\textbf{H3}] Participants will be more likely to respond that they are adhering to optimal assistance than to no assistance or object-highlighting assistance. 
    \item[\textbf{H3}] Participant agency will be unaffected by assistance fidelity. 
    \item[\textbf{H4}] Participants will perceive the AR system equipped with optimal assistance as more usable and useful than the AR system equipped with no assistance or object-highlighting assistance.
\end{itemize}

\subsubsection{Results summary} 
\begin{figure}[t]
    \centering
    \begin{minipage}[t]{.31\linewidth}
        \includegraphics[width=\linewidth]{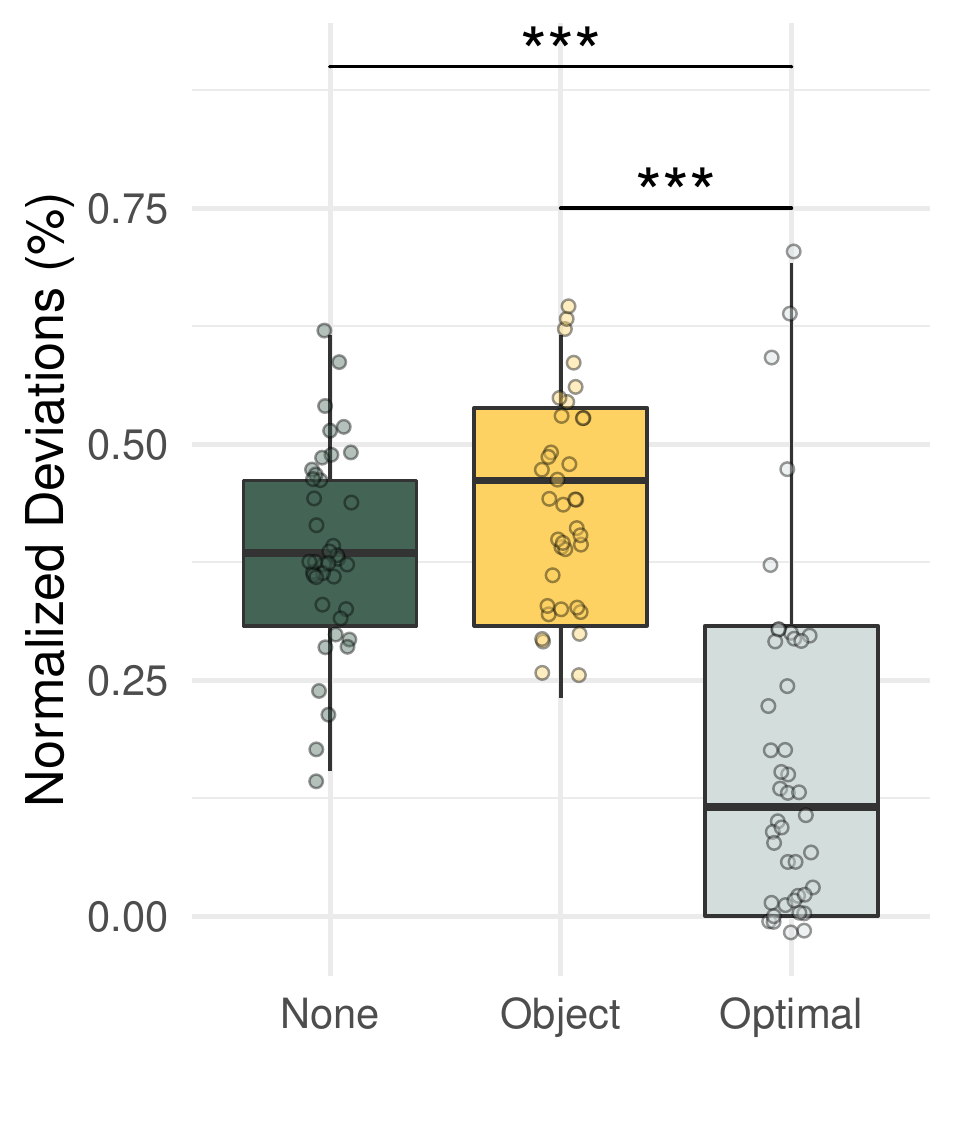}
        \caption{On average participants in the optimal condition deviated from the optimal ordering less frequently than participants in object-highlighting ($p<0.001$) and no assistance conditions ($p<0.001$).}\label{fig:easy_deviations}
    \end{minipage}\hfill
    \begin{minipage}[t]{.31\linewidth}%
        \includegraphics[width=\linewidth]{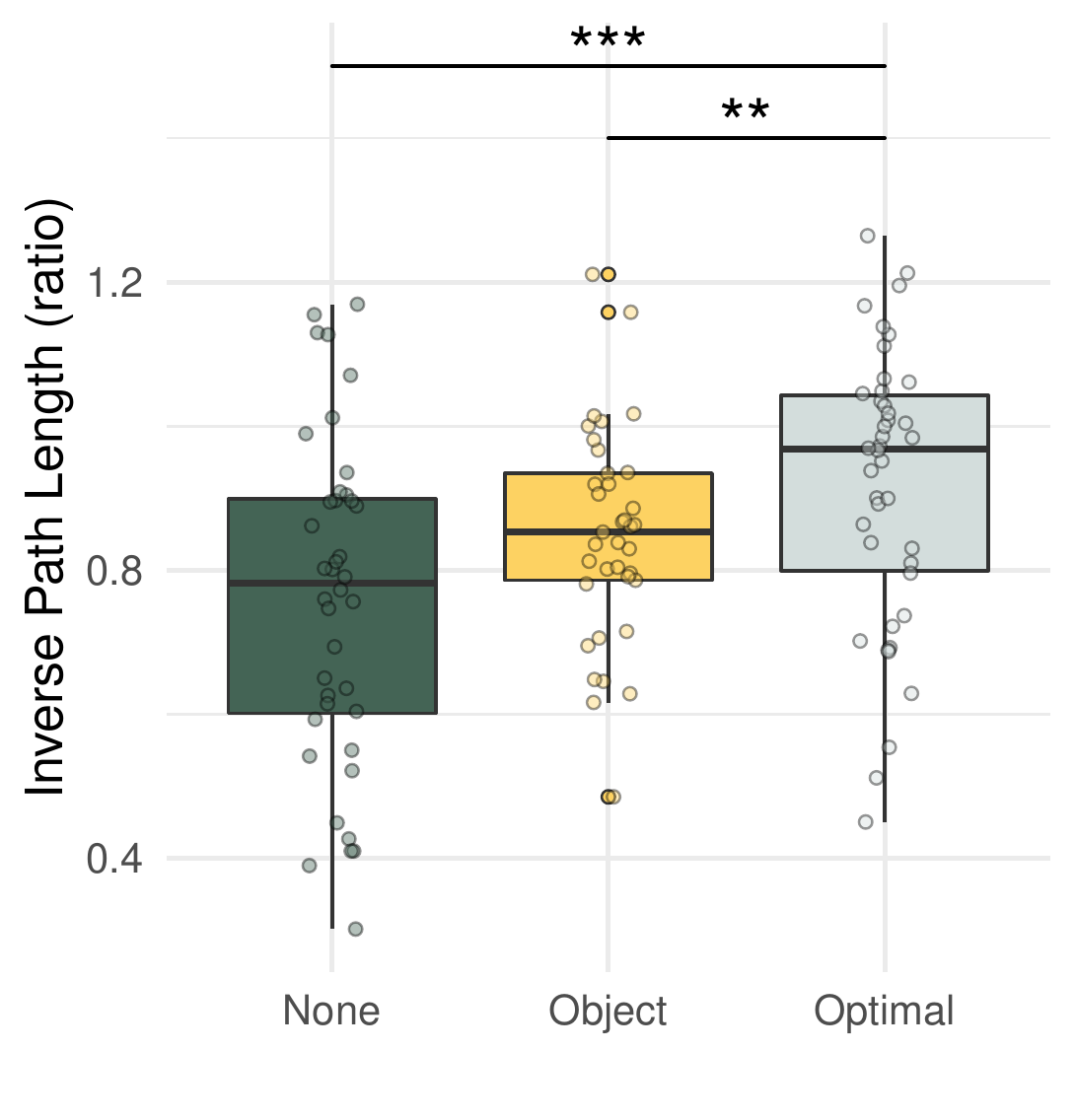}
        \caption{Participants provided with optimal assistance followed paths that more closely resembled the optimal paths than in object-highlighting ($p<0.01$) and no assistance conditions ($p<0.001$).}\label{fig:easy_ipl}
    \end{minipage}\hfill
    \begin{minipage}[t]{.31\linewidth}%
        \includegraphics[width=\linewidth]{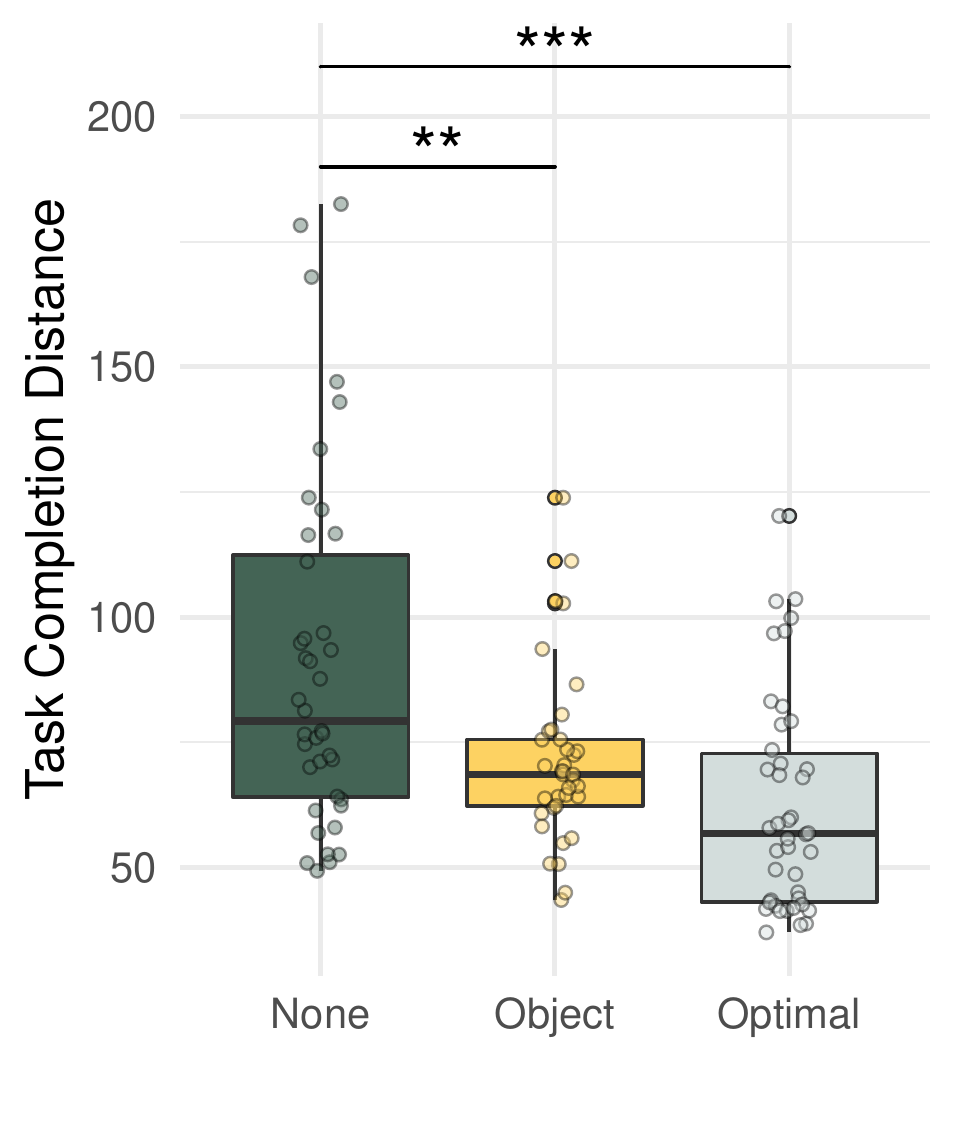}
        \caption{Participants provided with optimal assistance ($p<0.001$) and object-highlighting assistance ($p<0.01$) were more likely to find shorter paths than participants in the no assistance condition.}\label{fig:easy_distance}
    \end{minipage}
    \vspace{-0.15in}
\end{figure}

We now present results for the varying-assistance-fidelity study in terms of both objective and subjective metrics. For an in-depth reporting of the statistical analyses, see Sec.~\ref{sec:statAssistance} and Figs.~\ref{fig:easy_deviations}--\ref{fig:easy_distance}.

\textbf{Objective metrics}. In the case of optimal assistance, we observe that participants are more likely to pick and place items in the optimal order than they are in either the none or object-highlighting assistance conditions as measured by normalized deviations (Fig.~\ref{fig:easy_deviations}). \emph{This suggests that people may not able to compute independently the optimal ordering of location visits for object-rearrangement tasks based on only a first-person perspective, even with object highlighting}. Further, in the case of optimal assistance, participants follow the shortest-path trajectory between points more closely than they do in either the no assistance or object-highlighting assistance conditions (see IPL in Fig.~\ref{fig:easy_ipl}). \emph{This suggests that when participants freely navigate within an environment, they do not naturally take the shortest paths to get from point to point.} Taken together, these two findings support our hypothesis H1 that participants will follow optimal assistance when presented with it.

Even though participants tend to follow optimal assistance, this does not necessarily translate to improvements in all performance metrics. We measure task performance in two additional ways: total distance traveled and total task completion time. We observe that participants presented with object-highlighting and optimal assistance generated significantly shorter total paths than those generated in the no-assistance case, but the average path length traveled with these two forms of assistance was not substantially different (Fig.~\ref{fig:easy_distance}). \emph{This indicates that it is possible to get users to follow shorter paths than those they might find on their own, but that simply highlighting objects may be sufficient for decreasing user path distance. Interestingly, even though participants found shorter paths in the optimal and object-highlighting conditions, there was no significant difference in the total task completion time between the three conditions.} This further indicates that even though we are able to shorten path length, this may come at some cost to the speed at which a user completes the task (potentially in the interpretation of the interface). Ultimately, navigating this trade-off is likely user or task specific, and can likely be made more favorable with alternative and personalized interface design. Taken together, these two findings partially confirm our hypothesis H2 that participants will have higher task performance when presented with optimal assistance. 

\textbf{Subjective metrics}. \textit{Agency.} We found that participants generally felt in control of what they should do next and how quickly they completed the task. Even so, participants provided with optimal assistance, while still generally agreeing with feeling in control, reported that they felt less in control than in the other assistance conditions. Participants provided with no assistance and object-highlighting assistance were neutral about their feelings of needing to follow the assistance; participants provided with optimal assistance rated that they did feel the need to follow the assistance. Finally, participants provided with no assistance and object-highlighting assistance disagreed that they would prefer to have the system show them what to do next and where to go next. Since they were not exposed to a condition where they were provided this information, they are likely rating this against their idea of what such a system might look like. \emph{Participants who actually were exposed to optimal assistance agreed that they preferred this to not having this assistive information. So, even though participants in the optimal assistance condition felt less in control overall, they seemed to prefer this than to an alternative}. Overall, this does not support our hypothesis H3 that participants' sense of agency would remain unaffected by assistance fidelity; however, it seems that despite feeling a slight loss in their sense of agency, they may actually prefer this sacrifice in order to obtain useful information for optimal task completion.

\textit{Usability and Utility}. We found that participants generally felt that they were faster with assistance than without assistance. \emph{Participants provided with both optimal assistance and object-highlighting assistance believed that they completed the task faster with the additional information than they would have without it (i.e., with no assistance).} While participants in all assistance-fidelity conditions perceived the assistance to be useful, \emph{those provided with optimal assistance perceived it to be more useful than in either the object-highlighting or no assistance conditions.} These results are especially interesting given that no significant difference was found in total task completion time between assistance-fidelity levels. Overall, these findings partially support our hypothesis H4 that participants will be more likely to perceive the optimal assistance to be usable and useful than other assistance types. 

% \begin{figure}[t]
%     \centering
%     \begin{minipage}[t]{.31\linewidth}
%         \includegraphics[width=\linewidth]{figs/easy_deviations.pdf}
%         \caption{On average participants in the optimal condition deviated from the optimal ordering less frequently than participants in object-highlighting ($p<0.001$) and no assistance conditions ($p<0.001$).}\label{fig:easy_deviations}
%     \end{minipage}\hfill
%     \begin{minipage}[t]{.31\linewidth}%
%         \includegraphics[width=\linewidth]{figs/easy_ipl_end.pdf}
%         \caption{Participants provided with optimal assistance followed paths that more closely resembled the optimal paths than in object-highlighting ($p<0.01$) and no assistance conditions ($p<0.001$).}\label{fig:easy_ipl}
%     \end{minipage}\hfill
%     \begin{minipage}[t]{.31\linewidth}%
%         \includegraphics[width=\linewidth]{figs/easy_distance.pdf}
%         \caption{Participants provided with optimal assistance ($p<0.001$) and object-highlighting assistance ($p<0.01$) were more likely to find shorter paths than participants in the no assistance condition.}\label{fig:easy_distance}
%     \end{minipage}
% \end{figure}

\begin{figure}[t]
    \centering
    \includegraphics[width=.75\linewidth]{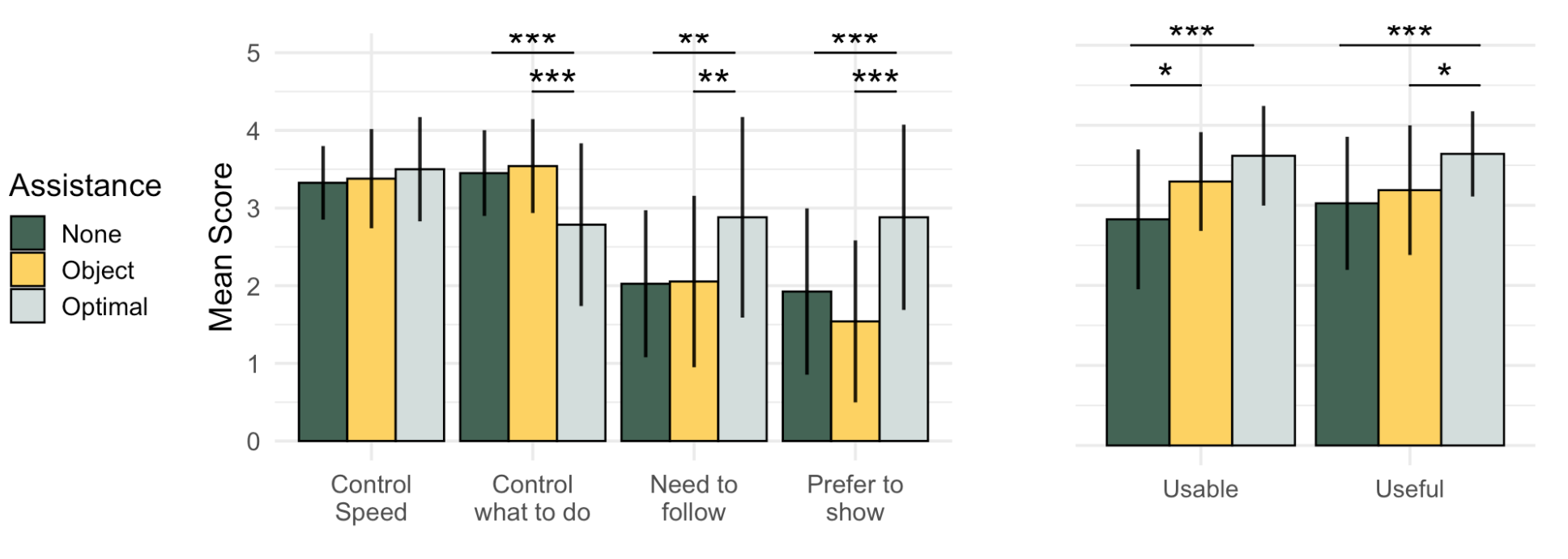}
        \caption{\textbf{Left}, When compared against no assistance and object-highlighting assistance, participants in the optimal assistance condition were less likely to feel in control (both $p<0.001$), more likely to feel the need to follow the assistance (both $p<0.01$), and more likely to prefer being told what to do (both $p<0.001$). \textbf{Right}, Participants in the optimal assistance were more likely to rate assistance as usable ($p<0.001$) and useful than those provided with no-assistance ($p<0.001$).  Those in object-highlighting more strongly agreed that the assistance was useful when compared to no-assistance ($p<0.05$). Those in optimal assistance more strongly agreed that the assistance was useful as compared to the object-highlighting assistance ($p<0.05$). }\label{fig:easy_monster}
   \vspace{-0.15in}
\end{figure}

\subsection{Interactions between assistance and task difficulty}\label{sec:interactions}

\subsubsection{Hypothesis.}
We had the following hypothesis:
\begin{itemize}
    \item[\textbf{H5}] As the task difficulty increases, participants will be more willing to accept the assistance.
\end{itemize}

\subsubsection{Results summary.}
\begin{figure}[t]
    \centering
    \begin{minipage}[t]{.06\linewidth}%
    \end{minipage}\hfill
    \begin{minipage}[t]{.4\linewidth}
        \includegraphics[width=\linewidth]{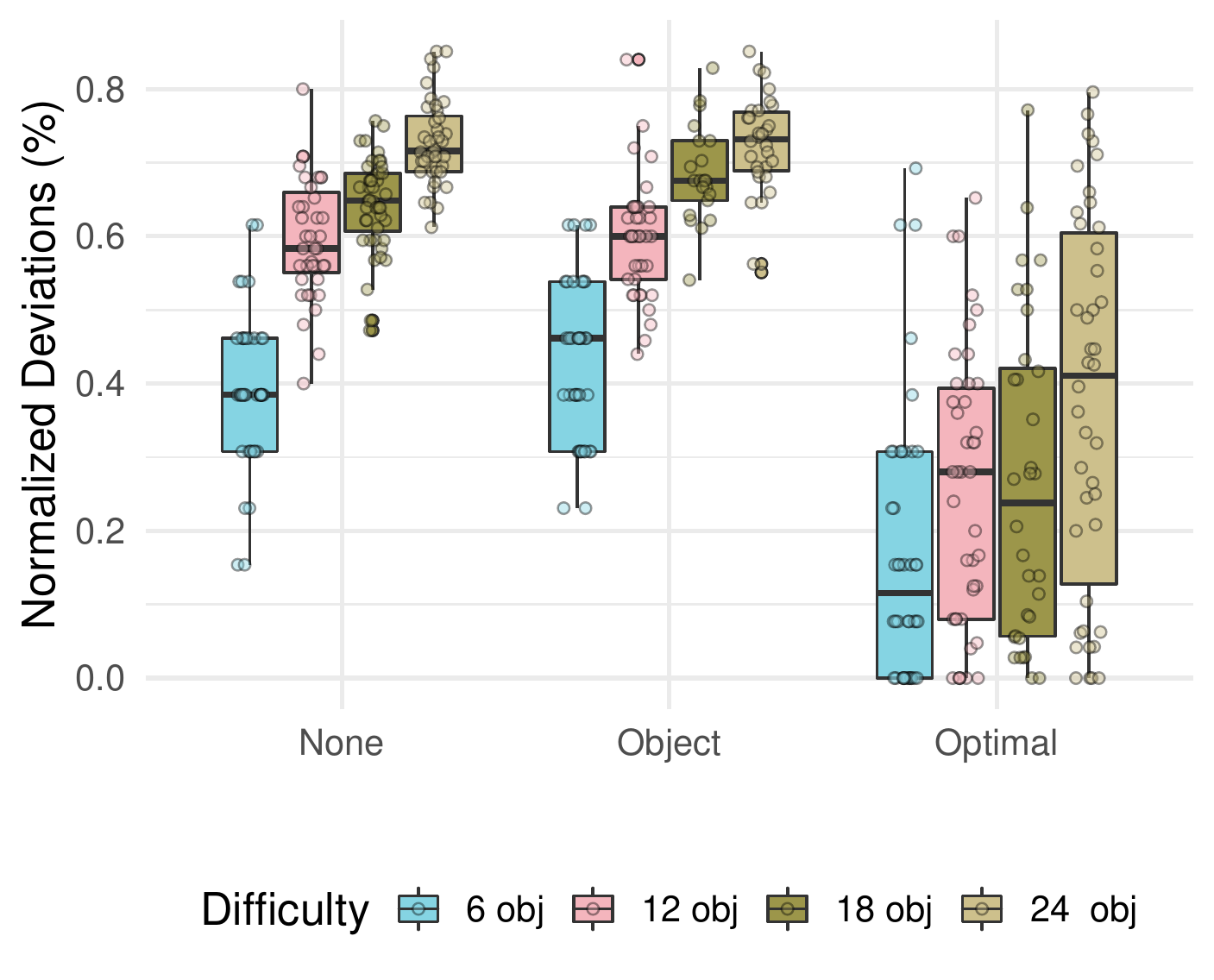}
        
    \end{minipage}\hfill
    \begin{minipage}[t]{.06\linewidth}%
    \end{minipage}\hfill
    \begin{minipage}[t]{.4\linewidth}%
        \includegraphics[width=\linewidth]{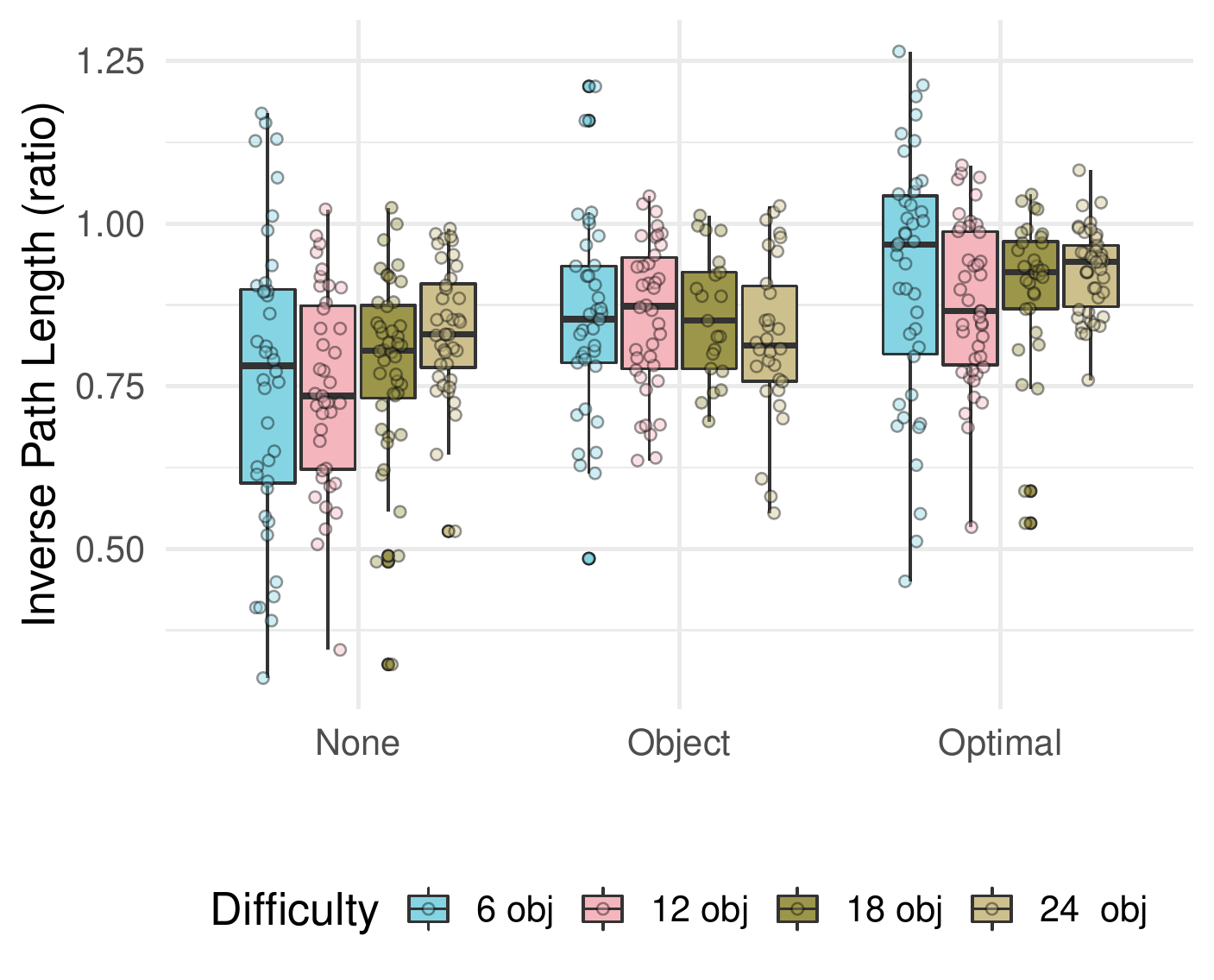}
      
    \end{minipage}\hfill
        \begin{minipage}[t]{.06\linewidth}%
    \end{minipage}\hfill
    \caption{\textbf{Left}, participants provided with optimal assistance were less likely to deviate from the optimal ordering than participants provided with either no assistance or object-highlighting assistance ($p<0.001$, both). As the difficulty increased, participants were more likely to deviate in each group. \textbf{Right}, participants provided with optimal assistance were more likely to follow shortest paths (higher IPL) than those provided with both object-highlighting ($p<0.01$) and no ($p<0.001$) assistance. Participants provided with object-highlighting assistance took shorter paths than those in none ($p<0.001$).}  \label{fig:int_ipl}
    \vspace{-0.15in}
\end{figure}

For an in-depth reporting of the statistical analyses, see Sec.~\ref{sec:statInteraction} and Fig.~\ref{fig:int_ipl}.

Across the board, we observed that participants deviated more from the optimal ordering as task-difficulty increased (Fig.~\ref{fig:int_ipl}). In the cases of no assistance and object-highlighting assistance, this trend conforms to our expectation that people have difficulties in finding the optimal solution in object-rearrangement tasks. We still observe this trend, however, when people are explicitly given the optimal ordering as in the optimal-assistance condition. Interestingly, though, the variance in the optimal-assistance condition is consistently much greater than those in either the no assistance or object-highlighting assistance cases. This suggests that the underlying distribution of assistance acceptance is likely multi-modal, and future work can investigate mechanisms to increase acceptance among various sub-populations of users. 

\emph{We also observed that participants were more likely to take shorter paths in the optimal-assistance condition than in the other two conditions, and that participants provided with object-highlighting assistance were more likely to take shorter paths than those provided with no assistance (Fig.~\ref{fig:int_ipl})}. However, there was no change in IPL as the task-difficulty level increased. Taken together, this refutes our hypothesis H5 that users would be more likely to accept assistance as the task became harder. In fact, the data suggest that they are either less likely or just as likely to accept the assistance as the task difficulty increases. 

\section{Conclusions}

This work has presented (1) a novel framework for computing and displaying AR assistance for object-rearrangement tasks that characterize a broad category of quotidian tasks, (2) a novel AR simulator that can enable web-based evaluation of AR-like assistance and large-scale data collection, and (3) a study that assesses how users respond to the proposed AR assistance in the AR simulator on a specific object-rearrangement task: house cleaning.

The study illustrated several salient trends. First, by following the optimal assistance, participants were able to reduce the overall distance they travelled, suggesting that people do not immediately solve this problem optimally and could benefit from a system like the one we propose. Second, though participants' reported feeling less in control over their own actions when following the optimal assistance they were also more likely to agree that they preferred a system that told them what to do than users in either other group. This indicates that users may be willing to sacrifice a small amount of agency in favor of a system that provides useful assistance. Finally, users were less likely or equally likely to accept the assistance as the difficulty of the task increased, though the population of users in the optimal assistance condition exhibited a much wider variability of assistance acceptance. This indicates that there are potential subgroups in the user population, and that future work should be conducted to discover these groups and develop personalized assistance systems.  

Future work will explore extensions of the current framework and study, including developing and assessing perception-based learned policies that assume lower fidelity of the embodied agent’s observations, considering multi-agent formulations that incorporate models of the user’s behavior, and extending the current assistance framework and AR simulator to other quotidian object-rearrangement tasks and assessing the user experience in those settings.

\begin{acks}
The authors would like to thank Gideon Stocek and Blaise Ritchie for setting up the back-end servers, front-end development and design of the study logic and flow, integration with ORTools for online optimal path calculations, and integration with psiTurk for deployment on AMT. The authors would also like to thank Yan Xu and Mei Gao for help with the UX design decisions, survey design, and interpretation of early results; Joshua Walton for feedback on the design of the assistance and HUD; Hrvoje Benko and Tanya Jonker for early feedback on study design; Michael Shvartsman for pointer to psiTurk and James Hillis for forming cross-function connection with the Habitat team. Lastly, the authors are grateful to Amanpreet Singh, Mandeep Baines, Oleksandr Maksymets, Alexander Clegg, and Dhruv Batra from the Habitat team for their support in understanding and debugging various features related to Habitat for our setup.
\end{acks}

\bibliographystyle{ACM-Reference-Format}
\bibliography{kondo}

\appendix

%\section{HIT description}\label{sec:studyprompt}

\section{Participants statistics in our study}\label{sec:studyparticipants}
Table \ref{tab:participants} shows the number of participant in each of our twelve conditions. Participants were uniformly randomly assigned to one of these conditions when they accepted to participate in the study. 

\begin{table}[h]
\centering
\begin{tabular}{|l|c|c|c|c|}
\hline
      & 6 objects & 12 objects & 18 objects & 24 objects\\ \hline
None  & 40 & 39 & 44 & 40  \\ \hline
Object-highlighting  & 37 & 38 & 21 & 30    \\ \hline
Optimal & 42 & 42 & 32 & 42     \\ \hline
\end{tabular}
\caption{The number of participants across conditions in our study.}
\label{tab:participants}
\end{table}

\section{Web application implementation details}
\label{sec:system_implementation}
The setup works as follows: we serve a Human Intelligence Task (HIT) to AMT using a psiTurk server, which also allows us to advertise our HIT on through the AMT web portal. Through this portal, participants can view a study description, compensation details, and the estimated completion time. After accepting a HIT, participants consent to study participation, which initiates serving the approximately 8 GB Habitat WebGL application to the user’s web browser using a combination of a psiTurk server and an NGINX server~\cite{reese2008nginx}. The majority of the application is loaded directly onto the participant’s computer, with the exception of the OR-Tools replanning module. When a disagreement between the participant’s actual path and the computed optimal path occurs, the participant’s web browser communicates with the psiTurk server to recalculate the optimal path and send this back to the client. After completing the survey, the data collected during the experiment (e.g., keyboard actions, time spent completing each phase, survey responses) are transmitted back to the server where it is stored in a MySQL database for later use. After the participant completes the experiment, we employ psiTurk to approve and disburse payment through AMT.

\section{Disorganized house generation}\label{sec:houseGeneration}

% \BN{Include a description and image of the dataset (both objects/bins and replica), how the receptacles were placed, and how the objects were placed.} 

We now describe how we generated the disorganized house for the study described in Section \ref{sec:task}.
\subsection{Bin placement}
Bins were placed manually within the environment in semantically reasonable locations. For example, the dish bin was placed on top of the counter next to the sink and the office supply box was placed on top of thew desk in the office. Receptacle locations were kept fixed in each difficulty setting. 

% To confirm that these receptacles were placed appropriately, we asked participants to rank their agreement to this statement: “” on a 5 point scale from Strongly Disagree to Strongly Agree. These results are discussed in Section \ref{sec:simulator_realism}.

% \BN{TODO:: VERIFY THAT THIS WAS REASONABLE WITH THE LIKERT DATA (or could put this in the appendix).}

\subsection{Object placement}
To sample object locations within the scene we randomly sampled a total of 40 navigable scene points using the Habitat simulator. We used rejection sampling in order to ensure that sampled points were at least one unit of distance away from every other sampled point and receptacle location within the scene. We then used the first N points from this list to define the object locations in the scene. For example, the lowest difficulty setting (6 objects) contained the first 6 points from this list, the highest difficulty setting contained the first 30 points. This way, each scene built on top of the previous scene in order to control for any single scene having an outlying dispersion between points. Each point was assigned a semantic object category, as well. This was kept consistent throughout each difficulty setting. The actual object model used for any individual point was held constant within difficulty settings, but was randomly sampled across difficulty settings. Thus, if a sampled point was  assigned to the books and magazines category in the lowest difficulty setting, it would be assigned to the books and magazines category in the highest difficulty setting, as well, but it might not be exactly the same book. Object locations for each difficulty setting are shown as circles in Fig.~\ref{fig:optimal_solution}.  

\subsection{Starting Location}
The starting position of each participant was held constant for any individual scene. This position was calculated by first finding the centroid of all of objects placed in the scene. This point was determined to be navigable using the Habitat simulator. If the point was not navigable, rejection sampling was used to find a navigable point within a small radius surrounding the centroid. This method allowed us to ensure that the participant’s starting position did not bias their solution by their starting location. Additionally, this method allowed us to generate our optimal assistance, discussed in Sec.~\ref{sec:CVRP}.

\section{Subjective metrics}\label{sec:fullsubjective}
The full list of Likert-scale questions used in our study is below (a subset was described in Sec.~\ref{sec:metrics} and used for analysis in Sec.~\ref{sec:assistance}):
\begin{enumerate}
    \item \textit{Agency}:
    \begin{itemize}
        \item[] ``I am in charge of deciding what step I complete next during the house cleaning task'' (\textit{Control what to do})
        \item[]``I am in charge of deciding where I go next during the house cleaning task''(\textit{Control where to go)})
        \item[]``I am responsible for the speed at which I completed the task'' (\textit{Control of speed})
        \item[]``I feel that I need to follow the suggestions given to me by the system'' (\textit{Need to follow})
        \item[] ``I prefer that the system show me what to do next rather than figure it out myself during the house cleaning task'' (\textit{Prefer to show what})
        \item[] ``I prefer that the system show me where to go next rather than figure it out myself during the house cleaning task'' (\textit{Prefer to show where})
    \end{itemize}
    \item \textit{Utility and Self-efficacy}:
    \begin{itemize}
        \item[] ``I completed the task as quickly as I could'' 
        \item[] ``I followed the help given to me by the system''
        \item[] ``I took longer than I needed to to complete the task''
        \item[]``I found the help given to me by the system to be useful'' \item[] ``I found a better way to complete the task than offered to me by the system''
    \end{itemize}
    \item \textit{Usability}:
    \begin{itemize}
        \item[]``I found the house cleaning task more difficult to complete than the training task''
        \item[]``I thought the system was easy to use''
        \item[]``I would imagine that most people would learn to use the help provided by the system very quickly''
        \item[]``I would like to use the help provided to me by system during the house cleaning task frequently in real life house cleaning scenarios''
        \item[]``The assistance provided to me by the system during the house cleaning task helped me complete the task faster than if I had used the help provided to me during the training task''
        \item[]``The help provided to me by the system during the house cleaning task was sufficient in order to accomplish the task''.
    \end{itemize}
\end{enumerate}

\section{Statistical analysis}
Secs.\ \ref{sec:statAssistance} and \ref{sec:statInteraction} provide statistical analysis for the studies performed in Secs.\ \ref{sec:assistance} and \ref{sec:interactions}, respectively.
\subsection{Effects of varying assistance fidelity}\label{sec:statAssistance}

\textbf{Objective metrics}. We performed four one-way ANOVA tests to measure the effect of assistance on normalized deviations, IPL, task completion distance, and task completion time. The effect of assistance was statistically significant on normalized deviations ($F(2,116) = 41.48$, $p=0.001$), IPL ($F(2,116) = 15.39$, $p=0.000$) and task-completion distance ($F(2,116) = 13.05$, $p=0.000$) (see Figs.~\ref{fig:easy_deviations}--\ref{fig:easy_distance}). No statistically significant effect of assistance was found on task-completion time ($F(2,116) = 0.652$, $p=0.523$). We performed post hoc Tukey honest significant difference tests to measure differences between each group. 

\textbf{Subjective metrics}.
We performed six one-way ANOVA tests to test for a main effect of assistance type within each question (Sec.~\ref{sec:metrics}). We found a statistically significant effect of assistance on Control what to do ($F(2,116) = 11.38$, $p=0.000$), Need to follow ($F(2,116) = 7.59$, $p=0.000$), Prefer to show ($F(2,116) = 15.56$, $p=0.000$), Useful ($F(2,116)=7.83$, $p=0.000$) and Usable ($F(2,116) = 12.72$, $p=0.000$). Following each ANOVA, we performed a post hoc Tukey test when a main effect was found; Fig.\ \ref{fig:easy_monster} summarizes these results. 

\subsection{Interactions between  assistance and task difficulty}\label{sec:statInteraction}

We conducted two individual two-way factorial ANOVA tests to measure the main effects of assistance type and task difficulty on normalized deviations and IPL, and any combined effects of assistance type and difficulty (see Fig.~\ref{fig:int_ipl}). Assistance type ($F(2,435) = 265.75$, $p=0.000$) and difficulty ($F(3,435) = 76.62$, $p=0.000$) had statistically significant effects on normalized deviations. The interaction between the two independent variables was also statistically significant ($F(6,435) = 2.865$, $p=0.001$). The effect of assistance type on IPL was statistically significant ($F(2,435) = 30.71$, $p=0.000$). There was no significant effect of task difficulty on IPL, and the interaction effect was not statistically significant. We performed post hoc Tukey honest significant difference tests, where effects were found.
% \section{Effects of Varying Task Difficulty}

% \subsection{Hypotheses}
% We had the following hypotheses:
% \begin{itemize}
%     \item[\textbf{AH1}] Increasing the ratio of objects to bins will make the task more difficult for participants to complete.
%     \item[\textbf{AH2}] Participants will perceive a higher ratio of objects to bins as more difficult than a lower ratio of objects to bins.
% \end{itemize}

% \section{Interactions between Assistance and Task Difficulty}

% \subsection{Additional Hypotheses}
% We had the following hypotheses:
% \begin{itemize}
%         \item[\textbf{AH3}] As the task difficulty increases, the benefits of optimal assistance will become more apparent.
% \end{itemize}

\end{document}